\documentclass[10pt,preprint]{emulateapj}

\newcounter{minirefcount}
\usepackage{amssymb}
\usepackage{amsmath}
\usepackage{wasysym}

\shorttitle{ALMA observations of AS 205}
\shortauthors{Salyk et al.}

\begin{document}

\title{ALMA observations of the T Tauri binary system AS 205: Evidence for molecular winds and/or binary interactions}

\author{Colette Salyk}
\affil{National Optical Astronomy Observatory, 950 N Cherry Ave, Tucson, AZ 85719, USA}
\author{Klaus Pontoppidan}
\affil{Space Telescope Science Institute, 3700 San Martin Drive, Baltimore, MD 21218}
\author{Stuartt Corder}
\affil{Joint ALMA Observatory, Av. Alonso de C\'{o}rdova 3107, Vitacura, Santiago, Chile}
\author{Diego Mu\~{n}oz}
\affil{Center for Space Research, Department of Astronomy, Cornell University, Ithaca, NY 14853, USA}
\author{Ke Zhang}
\affil{Division of Geological \& Planetary Sciences, Mail Code 150-21, California Institute of Technology, Pasadena, CA 91125, USA}
\author{Geoffrey A.\ Blake}
\affil{Division of Geological \& Planetary Sciences, Mail Code 150-21, California Institute of Technology, Pasadena, CA 91125, USA}

\begin{abstract}
In this study, we present high-resolution millimeter observations of the dust and gas disk of the T Tauri star AS 205 N and its companion, AS 205 S, obtained with the Atacama Large Millimeter Array.  The gas disk around AS 205 N, for which infrared emission spectroscopy demonstrates significant deviations from Keplerian motion that has been interpreted as evidence for a disk wind \citep{Pontoppidan11, Bast11}, also displays significant deviations from Keplerian disk emission in the observations presented here.  Detections near both AS 205 N and S are obtained in 1.3$\,$mm continuum, $^{12}$CO 2--1, $^{13}$CO 2--1 and C$^{18}$O 2--1.  The $^{12}$CO emission is extended up to $\sim2''$ from AS 205N, and both $^{12}$CO and $^{13}$CO display deviations from Keplerian rotation at all angular scales.  Two possible explanations for these observations hold up best to close scrutiny --- tidal interaction with AS 205 S or disk winds (or a combination of the two), and we discuss these possibilities in some detail.
\end{abstract}

\section{INTRODUCTION}

Observations of protoplanetary disks --- the birthplace of planets --- provide a window into the processes by which planets form.  While the dust component of protoplanetary disks has historically been easier to observe, the gas component contains the bulk of the disk mass, and encodes much of the important disk physics and chemistry.  Therefore, obtaining observations of the disk gas remains an important goal, in spite of its challenges.  To date, the majority of molecular gas detections in disks has come from infrared (IR) observations, where the heated disk atmospheres in the inner parts of disks (typically within $\sim$5 AU) produce emission from rovibrational or high-energy pure rotational molecular transitions \citep[e.g.][]{Najita03, Pontoppidan10b}.  Considerably fewer observations have been obtained of disk gas from the outer parts of disks, especially observations of the lowest-energy pure rotational transitions at millimeter wavelengths, as the sensitivity of current instruments allowed for the high-resolution study of only the brightest protoplanetary disks  \citep[e.g.][]{Dutrey97, Simon00, Oberg10}.   Yet it is the combination of observations across a wide range of wavelengths, probing a wide range of disk radii, that will enable a more complete understanding of disks.

This situation is about to change dramatically, as the superior sensitivity and image quality provided by the Atacama Large Millimeter Array (ALMA) is enabling extensive observations of disk gas at millimeter wavelengths.  In particular, ALMA observations will allow for detailed studies of its motion and distribution.  Indeed, ALMA observations have already revealed gaseous flows across a disk cavity \citep{Casassus13}, spirals \citep{Christiaens14} and a warped velocity field, consistent with fast radial inflow \citep{Rosenfeld14}, in the transitional disk HD 142527, as well as an inner clearing in the transitional disk Ophiuchus IRS 48 \citep{Bruderer14}.  ALMA will also allow for the study of winds and outflows from disks, probing more tenuous and smaller-scale regions than ever before; 
a helical disk wind has already been detected around the HD 163296 gas disk \citep{Klaassen13}.

The work presented here was inspired by an observational puzzle derived from IR observations.  For all moderate or high inclination disks, the Keplerian motion of the gas molecules is expected to produce double-peaked emission line profiles for the rovibrational lines produced within a few AU of the central star.  Surprisingly, however, observations have revealed that CO rovibrational emission lines from some moderate-inclination disks are strongly single-peaked (as defined by the so-called line profile parameter; \citealt{Bast11}).  As they also have a small spatial extent, these emission lines are inconsistent with any Keplerian disk models.  Furthermore, spectro-astrometry reveals that the lines cannot be explained solely as emission from an azimuthally-symmetric Keplerian disk \citep{Pontoppidan11}.  As a group, the disks with strongly single-peaked emission lines also tend to have high rovibrational CO line/continuum ratios, and signatures of high accretion rates (\citeauthor{Bast11}), including high UV fluxes \citep{Yang12} and high optical and IR veiling \citep[e.g.][]{Gahm08, Salyk08}.  In addition, several disks with single-peaked emission lines are known to have jets or outflows \citep{Alencar01, Wang04, Herczeg05}.  Strongly single-peaked profiles are seen in perhaps $\sim$30\% of CO rovibrational line-emitting disks (\citeauthor{Bast11}), but recent results suggest that nearly all protoplanetary disks produce emission lines that are more single-peaked than expected for a Keplerian disk, and that they tend to have a slight blue asymmetry \citep{Brown13}.  Therefore, whatever phenomenon explains these observations may be widespread.

A possible explanation for these surprising observations was suggested by \citet{Pontoppidan11}, who showed that disk + disk wind models could reproduce the observed lineshapes, spectro-astrometric line profiles and small spatial extent.  However, while the models used by \citeauthor{Pontoppidan11} were originally derived from magneto-hydrodynamic wind simulations \citep{Kurosawa06}, they were largely parameterized, and only loosely constrained by the available data.  Yet, if the disk wind models used by \citeauthor{Pontoppidan11} are correct, disks may be losing mass at a rate of $\sim$ 1\% of their mass accretion rate, and the results of \citet{Brown13} suggest that disk winds may in fact be universal.  How might these winds affect disk evolution, or the formation of planets?  The answers to these questions are essentially unknown, and require that we gain a better understanding of the properties and driving mechanisms of these winds.   If the disk wind cools quickly as it extends outwards, and its density drops, a low-critical-density, low-excitation tracer like $^{12}$CO $J=2-1$ may be a good choice for detecting a wind on larger scales, and for studying its structure and kinematics.

In this work, we present ALMA observations of CO $J=2-1$ from a prototypical source displaying evidence for a possible disk wind.   The target of our study --- the T Tauri star AS 205 N --- shows strongly centrally peaked IR molecular emission lines with high line/continuum ratios \citep{Salyk08, Pontoppidan11}, a high and possibly highly variable accretion rate \citep{Eisner05, Salyk13}, and some evidence for an outflow \citep{Mundt84}.  AS 205 N is also the northern and more massive component of a $1.3''$ (projected separation) binary system, in which the secondary star (AS 205 S) is itself a double-lined spectroscopic binary \citep{Eisner05}.  AS 205 is located at the northern edge of the Ophiuchus star forming region, at a distance of $\sim$125 pc \citep{Mamajek08,Loinard08}.

In Section \ref{sec:observations}, we briefly describe the data acquisition and reduction.  In Section \ref{sec:results}, we present detections of $^{12}$CO, $^{13}$CO and C$^{18}$O from the disks in the AS 205 binary system, and discuss interesting features in the images and velocity fields.  Then, in Section \ref{sec:origin} we discuss possible explanations for these features, including an extensive discussion of their possible connection to the disk winds proposed to explain AS 205 N's unusual IR spectroscopic features.  Finally, we include a brief disussion of implications (Section \ref{sec:discussion}) and conclusions (Section \ref{sec:conclusions}).

\section{OBSERVATIONS AND REDUCTION}
\label{sec:observations}
Observations of AS 205 N  and S were obtained on UT 27 Mar 2012 and 04 May 2012, with the Cycle 0 extended configuration.   The nearby quasar J1625-254, at a distance of $7.6^\circ$, was observed at $\sim$11 minute intervals to correct for atmospheric variations. Short-term phase variations were corrected using the on-board water vapor radiometers.  Fifteen 12-meter antennas were available for both sets of observations, although one to two antennas were flagged at any given time during the observations.  On 27 Mar, the minimum and maximum baselines were 43 and 402 meters (corresponding to minimum and maximum spatial scales of $\sim0.7''$ and $4''$, 88 and 500 AU at 125 pc), respectively, and on 04 May, they were 21 and 402 meters (with a maximum spatial scale of $\sim8'' = 1000$ AU).  The total on-source integration time for each set of observations was 15.8 minutes, for a total integration time of 31.6 minutes.

Observations were obtained in four spectral windows of width 117 MHz, with a channel width of 30.5 kHz (0.04 km s$^{-1}$), corresponding to a spectral resolution of 61 kHz (0.08 km s$^{-1}$).  Three of the spectral windows were centered at the frequencies of $^{13}$CO $2-1$ (220.399 GHz), $^{12}$CO $2-1$ (230.538 GHz) and C$^{18}$O $2-1$ (219.56 GHz).  The fourth window was centered on an undetected CH$_3$OH transition (232.415 GHz), and was used for continuum measurement.

The data were reduced according to a standard procedure, using the Common Astronomy Software Applications package (CASA; \citealp{McMullin07}).  Reduction steps included passband calibration, correction for atmospheric variations, and absolute flux calibration.  Absolute flux was calibrated using observations and models of Titan. The calibrated measurement sets were then combined for the two dates.  The continuum window was binned to a single channel, then CLEANED using the Clark algorithm and Briggs weighting with a robustness parameter of 0.5. The spectral windows were first continuum-subtracted, then binned to a channel spacing equal to the spectral resolution (0.08 km s$^{-1}$) and CLEANed in the same manner as the continuum window.  Continuum subtraction was performed with the CASA routine {\it uvcontsub} using a fit to the 232 and 219 GHz windows and either a constant model (for $^{12}$CO) or a linear model (for $^{13}$CO and C$^{18}$O).  Inspection of line-free channels suggests continuum-subtraction residuals in the $^{12}$CO maps of up to $\sim$10\% near the centers of AS 205 N and S; no significant residuals remain in the 
$^{13}$CO and C$^{18}$O maps.

\section{RESULTS }
\label{sec:results}
\subsection{Basic results}
\label{sec:basic}
\begin{figure*}
\includegraphics*[scale=0.8]{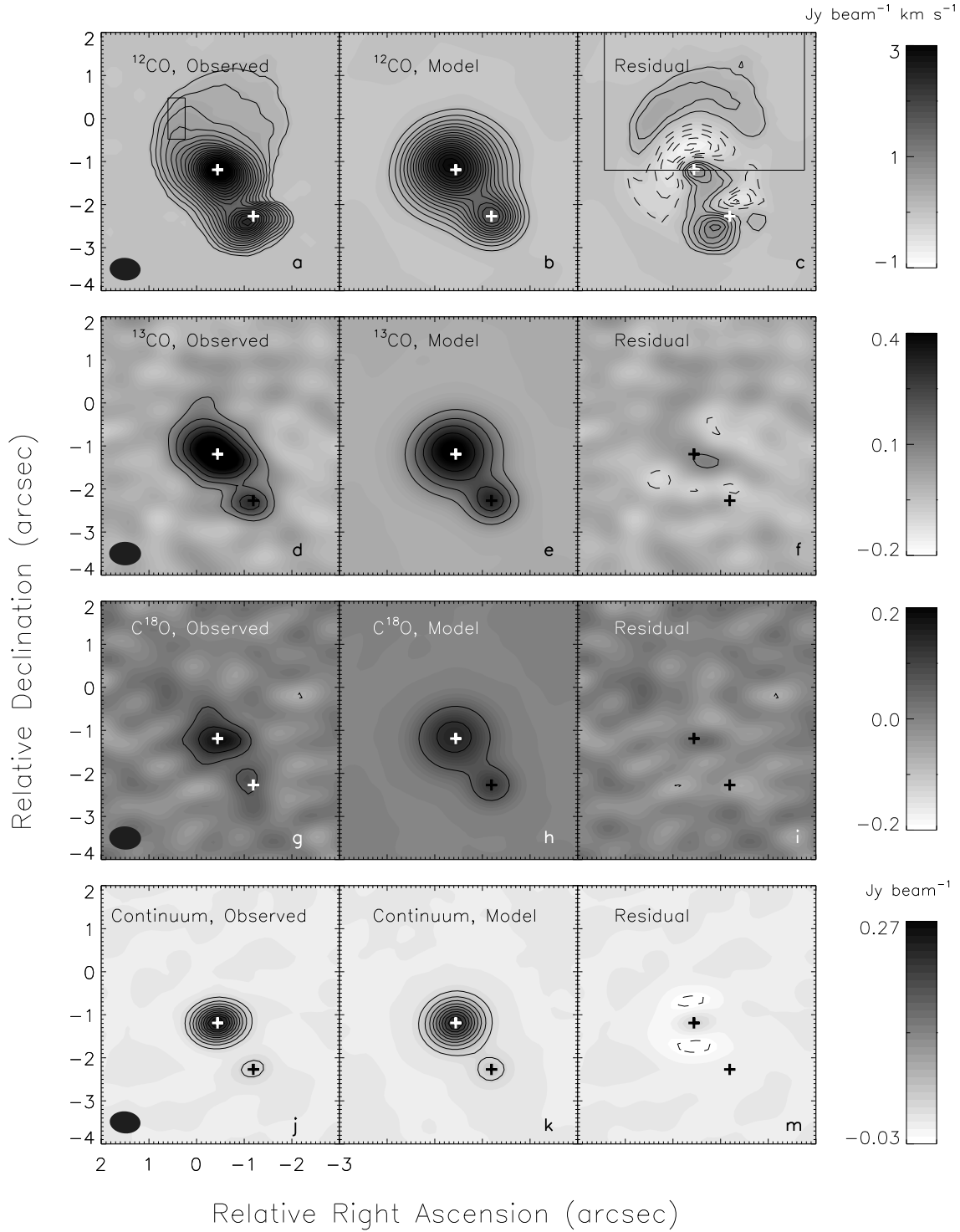}
\caption{a) Observed $^{12}$CO $2-1$ moment zero (intensity) map.  The box is used for calculations in Section \ref{sec:mass}; b) Modeled $^{12}$CO $2-1$ moment zero map; c)  $^{12}$CO $2-1$ model residuals.  The box is used for calculations in Section \ref{sec:mass}; 
d) Observed $^{13}$CO $2-1$ moment zero map; e) Modeled $^{13}$CO $2-1$ moment zero map; f)  $^{13}$CO $2-1$ model residuals; 
g) Observed C$^{18}$O $2-1$ moment zero map; h) Modeled C$^{18}$O $2-1$ moment zero map; i) C $^{18}$O $2-1$ model residuals; 
j) Observed continuum intensity; k) Modeled continuum intensity; m) Continuum model residuals.  Note the different grayscales for each set of maps. For the $^{12}$CO $2-1$ and continuum observations, contours begin at 10$\sigma$ and proceed in 10$\sigma$ steps.  For $^{13}$CO and C$^{18}$O, contours begin at 3$\sigma$ and proceed in 3$\sigma$ steps.  Dashed contours are negative.  In all panels, the crosses mark the measured center of the continuum emission from AS 205 N and AS 205 S. The beam ellipse is shown in the bottom of the left-most panels.
\label{fig:map_plot}}
\end{figure*}

In Figure \ref{fig:map_plot}, we show the observed (cleaned) $^{12}$CO $2-1$, $^{13}$CO $2-1$ and C$^{18}$O $2-1$ zeroth moment maps, and 1.3 mm continuum images, of the AS 205 system.  Note the different grayscales and contour intervals for each row.  Based on the continuum measurements, the separation of the two disks is $1.3''$, and the position angle of the binary is 214$^\circ$.  Previous works have found separations of $1.3''-1.4''$ and position angles between 204$^\circ$ and 213$^\circ$ \citep{Reipurth93, Ghez93, Koresko02, Prato03, Eisner05}, with no coherent temporal evolution.  Both sources appear to be detected in all three lines and in the continuum, and basic statistics for these detections are listed in Table \ref{table:statistics}.

For all maps, the synthesized beam is approximately $\sim 0.7'' \times 0.5''$ with a position angle of $\sim 90^\circ-100^\circ$ degrees.  The continuum emission appears unresolved or barely resolved, with a spatial FWHM ranging from $0.63''-0.76''$, and a position angle of 99$^\circ$ --- consistent with that of the synthesized beam.  The line emission is spatially extended in all three isotopologues, and for both components of the system, although the spatial extent of the northern component is larger than for the southern component.  The $^{12}$CO and $^{13}$CO emission from the northern component appear asymmetric, with a larger spatial extent towards the north and northwest.  

Although all isotopologues are present in the zeroth moment maps near AS 205 S as well as AS 205 N, the $^{13}$CO emission spatially associated with AS 205 S is actually dominated by gas at a similar velocity to AS 205 N. Therefore, some of this emission may instead be associated with AS 205 N.   We also note that the C$^{18}$O emission near AS 205 S is detected in the zeroth moment map only, and at only the 3$\sigma$ level.  Therefore, we caution that this detection is tentative, and could, for example, arise from errors in continuum subtraction.

\begin{figure*}[!ht]
\includegraphics[scale=0.6]{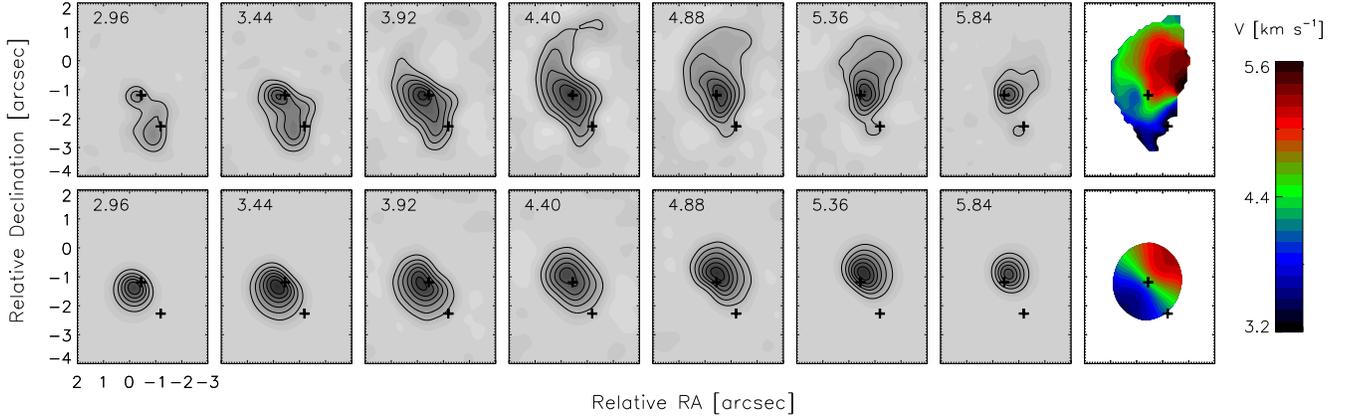}
\caption{Channel maps of observed $^{12}$CO emission (top) and a disk model (bottom), in addition to first moment maps (right), for AS 205 N.  Contours are multiples of 10$\sigma$.  Labels in upper left show the central velocity (v$_{LSR}$) of the channel.
\label{fig:channel_plot_12co}}
\end{figure*}

\begin{figure*}[!ht]
\includegraphics[scale=0.6]{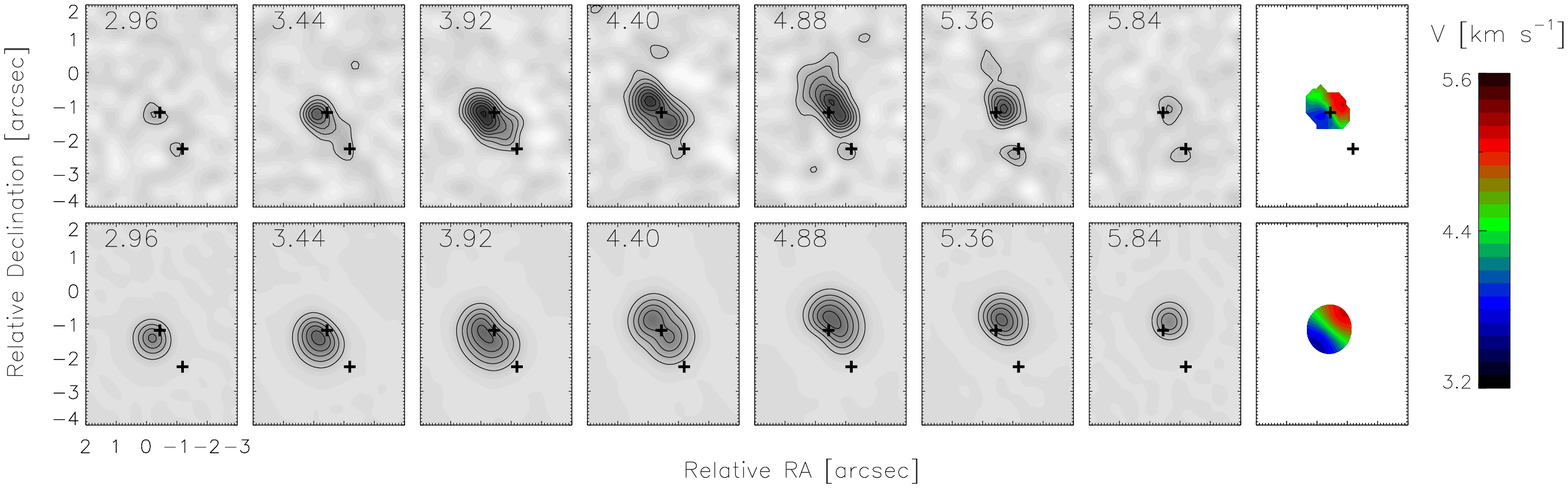}
\caption{Channel maps of observed $^{13}$CO emission (top) and a disk model (bottom), in addition to first moment maps, for AS 205 N.  Contours are multiples of 3$\sigma$.
\label{fig:channel_plot_13co}}
\end{figure*}

\begin{figure*}[!ht]
\includegraphics[scale=0.6]{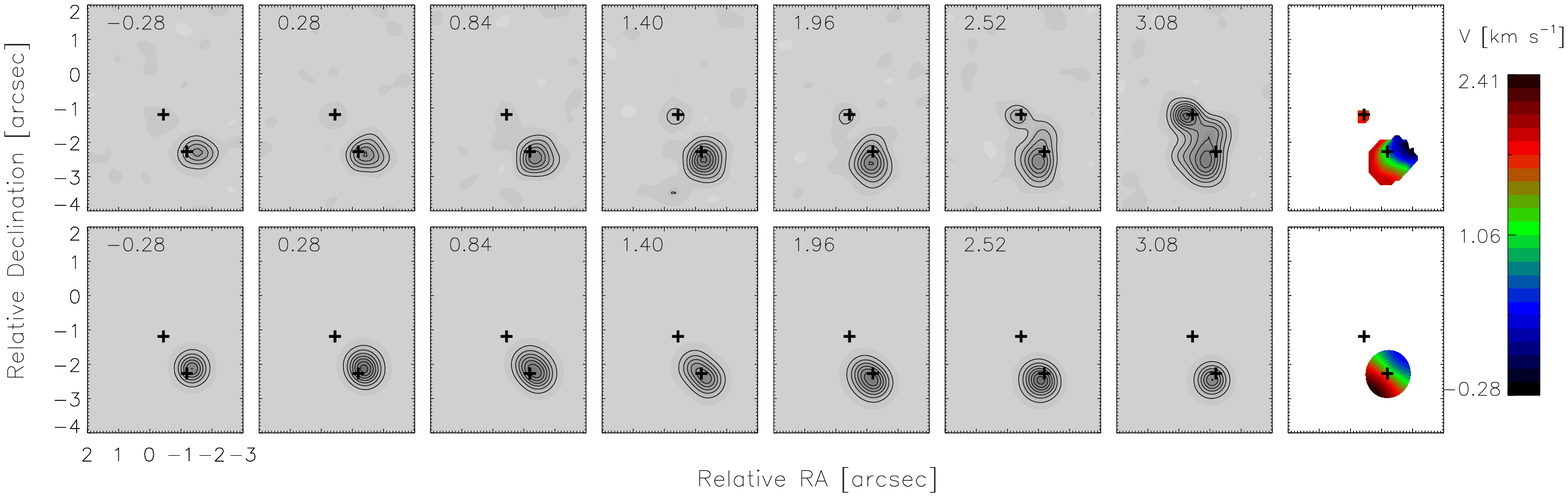}
\caption{Channel maps of observed $^{12}$CO emission (top) and a disk model (bottom), in addition to first moment maps, for AS 205 S.  Contours are multiples of 3$\sigma$.
\label{fig:channel_plot_12co_south}}
\end{figure*}

\begin{figure}
\includegraphics[scale=0.6]{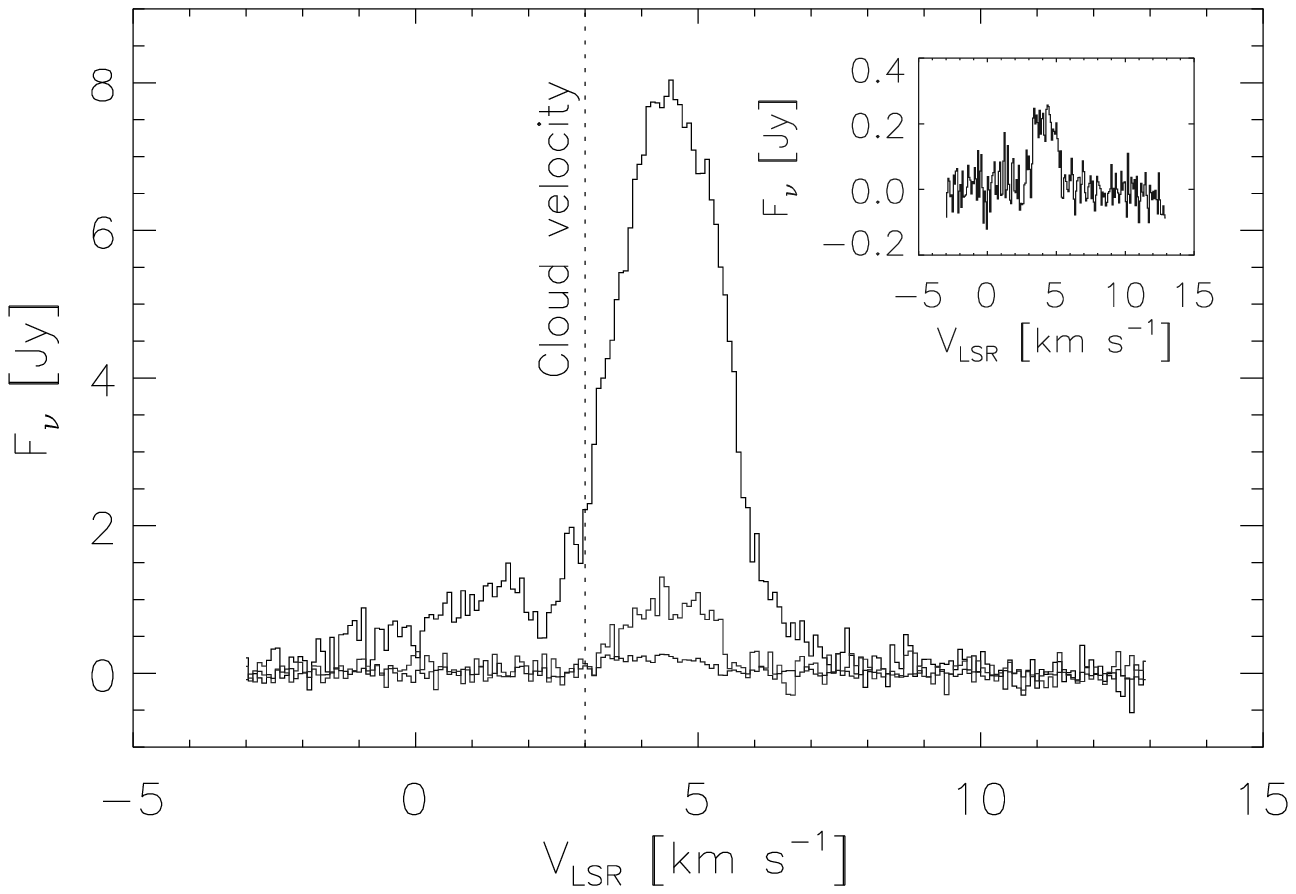}
\caption{$^{12}$CO (black), $^{13}$CO (light gray) and C$^{18}$O (dark gray, and inset) emission line profiles.  $^{12}$CO and $^{13}$CO line profiles are derived from the whole system.  The C$^{18}$O is from the northern component only, to reduce noise.   Excess $^{12}$CO emission at blue wavelengths is contributed by the southern component.  The vertical dotted line marks the measured cloud velocity near AS 205.
\label{fig:lineprofile_plot}}
\end{figure}

Channel maps and first moment (velocity field) maps for the northern component in $^{12}$CO and $^{13}$CO are shown in Figures \ref{fig:channel_plot_12co} and \ref{fig:channel_plot_13co}.  Channel maps and first moment maps for the southern component are shown in $^{12}$CO in Figure \ref{fig:channel_plot_12co_south}. Line profiles for the three isotopologues are shown in Figure \ref{fig:lineprofile_plot}. $^{12}$CO and $^{13}$CO lines are computed for the entire system, including the northern and southern components and any extended emission, using a $6.48''$ high by $4.32''$ wide extraction box.  The emission line for C$^{18}$O was computed for the northern component only, using a small $1.68''$ high by $1.44''$ wide extraction box, to reduce the contribution from the noisy background.  The fluxes and linewidths are summarized in Table \ref{table:statistics}.  The $^{12}$CO emission has a total flux (for both components combined) of 22.25 Jy km s$^{-1}$, consistent with the flux of $20.06\pm0.13$ Jy km s$^{-1}$ found by \citet{Oberg11a} if we allow for a 15\% error in absolute flux calibration.  

Using the mean and standard deviation of the line centers, we estimate $v_{LSR}$ of AS 205 N to be $ 4.29 \pm 0.18$, consistent to within 1.6 $\sigma$ of the published stellar velocity of $1.97 \pm 1.51$ from \citet{Melo03}, but inconsistent with the $-0.25\pm0.87$ km s$^{-1}$ found by \citet{Eisner05}.  The high and variable veiling in this source \citep{Gahm08} could contribute to large systematic errors in stellar velocities.  For AS 205 S, we take the $v_{LSR}$ for $^{12}$CO ($0.80\pm0.09$ km s$^{-1}$) to be the most likely value, since we were able to separate the signals from AS 205 N and AS 205 S (see Table \ref{table:statistics}).  \citet{Eisner05} find stellar velocities for the components of this spectroscopic binary of 
11.05$\pm$0.46 and $-6.36\pm1.11$ km s$^{-1}$, with an average of $2.35 \pm 1.2$ km s$^{-1}$, consistent with our result.

\begin{deluxetable*}{llll}
\tablecaption{ Detection statistics and line parameters
\label{table:statistics}}
\tablehead{\colhead{Tracer}&\colhead{Flux (N,S)}&\colhead{FWHM (N,S)} &\colhead{Line center (N,S)\tablenotemark{a}}\\
&\colhead{[Jy km s$^{-1}]$}& \colhead{[km s$^{-1}$] }& \colhead{[km s$^{-1}$]}}\\
\startdata
$^{12}$CO & $ 19.23\pm  0.27$, $  3.02\pm  0.22$ & $  2.27\pm  0.03$, $  2.59\pm
  0.18$ & $  4.48\pm  0.01$, $  0.80\pm  0.09$\\
$^{13}$CO & $  1.86\pm  0.04$, $  0.99\pm  0.09$ & $  1.62\pm  0.04$, $  2.68\pm
  0.24$& $  4.45\pm  0.02$, $  4.26\pm  0.12$\\
C$^{18}$O & $  0.34\pm  0.03$, $  0.13\pm 0.02$ & $  1.62\pm  0.14$  & $  4.18
\pm  0.07$ \\
\hline
&Flux density (N,S)\\
&[mJy]\\
\hline
continuum & $   377\pm     2$ , $    64\pm     1$ &\nodata&\nodata\\
\enddata

\tablenotetext{a}{$v_{LSR}$}
\tablenotetext{b}{It is not straightforward to separate the contributions from AS 205 N and S.  For these calculations, we perform a Gaussian fit
to the full $^{12}$CO emission profile to determine the AS 205 N statistics.  We then subtract the Gaussian fit from the line to obtain the
emission line from AS 205 S.}
\tablenotetext{c}{Here, we compute the statistics for AS 205 S by totaling the flux in the vicinity of AS 205 S.  For $^{13}$CO, the derived line center of $\sim$4 km s$^{-1}$ suggests significant contamination from AS 205 N.}
\end{deluxetable*}

\subsection{Disk Models}
\label{sec:models}
To aid with the analysis of the data, we constructed disk models to compare with the observed data.   Dust radiative transfer was performed with RADMC \citep{Dullemond04}, and line radiative transfer with RADLite \citep{Pontoppidan09}.  Simulated ALMA datasets were created using the CASA routines {\it simanalyze} and {\it simobserve}.  All simulated observations use the antenna configuration from the observations on 04 May 2012, and an exposure time of 30 minutes.  Simulated datasets are shown in Figures \ref{fig:map_plot} -- \ref{fig:channel_plot_12co_south}.  

Our nominal disk model for the northern component uses most of the same stellar and disk parameters derived by \citet{Andrews10} from a combined fit to the spectral energy distribution and millimeter visibilities --- $R_\star=3.7\,R_\odot$, $M_\star=1\, M_\odot$ and $T_\star=4250\,K$, $R_\mathrm{in}=0.14\, \mathrm{AU}$, $M_\mathrm{disk}=0.029\,M_\odot$ and $H\sim20\,\mathrm{AU}$ at 80 AU, with $H/R\propto (R/80\, \mathrm{AU})^{0.11}$. However, in contrast to \citet{Andrews10} who use the similarity solution for surface density, $\Sigma$, we use a broken power-law $\Sigma$,  $\propto (R/\mathrm{AU})^{-0.9}$ out to $R_\mathrm{out}=80$ AU, and $ \propto (R/\mathrm{AU})^{-12}$ beyond.  We also use a simple grain opacity with 80\% small amorphous astronomical silicates and 20\% carbon grains.  Less information is available in the literature for the southern component, and so for simplicity we utilize many of the same model parameters as for AS 205 N.  However, we assume $R_\star=4.2\,R_\odot$,  $M_\star=0.6\, M_\odot$ and $T_\star=3450\,K$, values derived by \citet{Prato03} but with a correction for the fact that the southern source is a spectroscopic binary \citep{Eisner05}.  We find a good match to the observed continuum data with a disk mass of $0.006\, M_\odot$.

The CO 2--1 emission is dominated by disk regions with densities well above the critical density for this line \citep[e.g.][]{Bruderer12}, and even the more tenuous winds we are investigating here are predicted to have densities at or above the critical density out to large distances from the star \citep[][]{Panoglou12}.  Therefore, all models assume local thermodynamic equilibrium conditions.  We assumed a gas to dust ratio of 100, and used a parameterized gas temperature increase in the disk atmosphere, following the procedure described in \citet{Zhang13}.  In particular, we set $T_\mathrm{gas}-T_\mathrm{dust}=300 \times e^{-r/30 AU}$ above the $\tau=1$ disk surface.  We also assumed an H$_2$/CO ratio of 5000 \citep{Lacy94} and a CO freeze-out temperature of 17 K.  For isotopologue abundances, we assumed $^{12}$CO/$^{13}$CO$=60$ and $^{12}$CO/C$^{18}$O$=500$.  Since our limited information would result in strong model degeneracies were we to try to exactly match the observed data, we simply made small adjustments to the total flux produced by the models to approximately match the data.  For AS 205 N, we multiplied the continuum models and the $^{12}$CO, $^{13}$CO and C$^{18}$O continuum-subtracted models by factors of 1.3, 1.4, 0.4 and 0.2.  For the southern source models, we multiplied by factors of 1, 1.2, 0.5 and 0.3.  None of the conclusions in this work depend on the absolute flux of the models.  

We find that an inclination of $\sim$15$^\circ\pm5^\circ$ is required to fit the observed $^{13}$CO velocity field for AS 205 N. Previous spectro-astrometric results, which also resolved the binary, found a disk inclination of $\sim$20$^\circ$.   Using the $^{12}$CO velocity field, we find a disk inclination of 25$^\circ\pm10^\circ$ for AS 205 S.  These inclinations and error bars are determined via by-eye examination of the channel maps and residuals, and based on a combination of the relative fluxes in the low and high-velocity channels, as well as the shapes of the maps.  We also find that the disk around AS 205 S must be significantly smaller than AS 205 N, and set $R_\mathrm{out}=35$ AU.  We discuss this result in more detail in Section \ref{section:truncation}.  There is some uncertainty in the literature about the mass of AS 205 S.  \citet{Eisner05} derive a larger mass than assumed in our nominal model, $1.28\, M_\odot$, along with temperatures for AS 205 N and S of 3800 and 4000 K, and a total luminosity of $0.7 L_\odot$.  With this alternative choice of parameters, we find an inclination for AS 205 S of $\sim15^\circ$.  Given the limited resolution of our observations compared to the size of the AS 205 S disk, spatial information cannot be used to break the degeneracy between stellar mass and inclination.

Figure \ref{fig:channel_plot_13co} shows that the AS 205 N disk model deviates from the $^{13}$CO data in some respects.  In particular, the high velocity emission is slightly over-predicted.  However, decreasing the inclination of the model would produce maps less elongated than the data at low velocities.   A similar discrepancy was noted by \citet{Rosenfeld12} for TW Hya, and attributed to a warp in the TW Hya disk.  In our case, the discrepancy could instead be related to the other evidence for non-Keplerian motions in our dataset, which is discussed in much greater detail throughout this work. 

The position angles of the AS 205 N and AS 205 S disks are observed to be $\sim 135^\circ \pm5^\circ$ and $\sim320^\circ \pm10^\circ$, respectively, and are therefore misaligned by nearly 180$^\circ$.  The relative alignments of the disk rotation axes cannot be unambiguously determined unless the near and far sides of both disks are known.  The disks are either oriented in the same direction with respect to the plane of the sky, with rotation vectors nearly anti-aligned, or they are oriented in opposite directions with respect to the plane of the sky, with rotation vectors having a misalignment of $40^\circ (25^\circ+15^\circ)$.

The $^{13}$CO emission from AS 205 N shows a distinct NE/SW asymmetry, with NE regions being brighter.  This is especially apparent in the 4.40 km s$^{-1}$ channel in Figure \ref{fig:channel_plot_13co}, for example.  If attributed solely to disk emission, this effect is naturally produced if the NE side is the far side of the disk, due to a combination of projection and optical depth effects \citep[e.g.][]{Pontoppidan09, Rosenfeld13}.  This orientation is consistent with the spectro-astrometric results of \citet{Pontoppidan11}.  However, this interpretation of the asymmetry is not definitive, as our dataset includes several features that are not well explained by a disk model alone.  It is not possible to determine the near/far-side orientation of the disk around AS 205 S with these data.

\subsection{Truncation of disk around AS 205 S}
\label{section:truncation}
\begin{figure}
\includegraphics[scale=0.5]{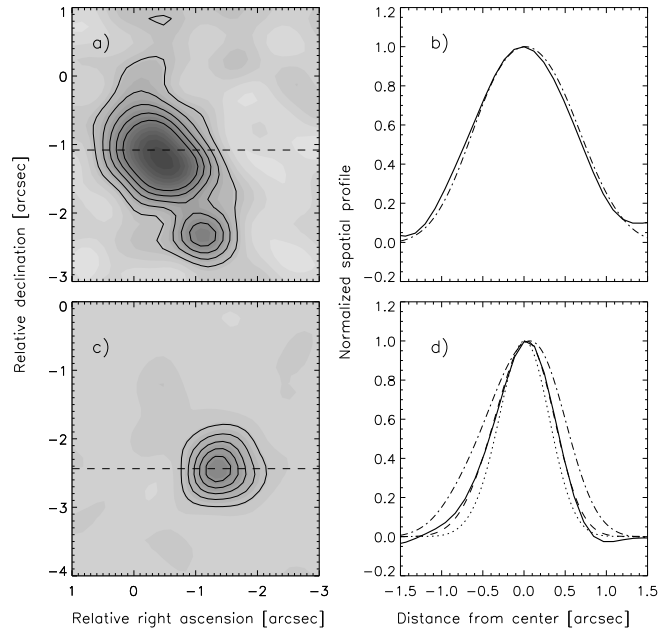}
\caption{a) Zeroth moment map of $^{13}$CO $2-1$ emission from AS 205 N.  The dashed line shows where the spatial profile is extracted.  b) The extracted spatial profile for AS 205 N derived from the data (solid line) and the nominal disk model with $R_\mathrm{out}=80$ AU (dot-dashed). c) Channel map of $^{12}$CO emission from AS 205 S at $v=0.84$ km s$^{-1}$ (same as the 3rd panel in the top half of Figure \ref{fig:channel_plot_12co_south}).  The dashed line shows where the spatial profile is extracted.  d) The extracted spatial profiles for AS 205 S.  The solid line shows the data, while the dot-dashed, dashed, and dotted lines show disk models with $R_\mathrm{out}=60$ AU, 35 AU and 20 AU.
\label{fig:spatial_profile_combined}}
\end{figure}

In this work, our disk models assume a broken power-law surface density that declines sharply at some radius $R_\mathrm{out}$.  For AS 205 N, the data are consistent with $R_\mathrm{out}\sim80$ AU.  This is demonstrated in Figure \ref{fig:spatial_profile_combined}, which shows a normalized spatial profile along a cut in the zeroth moment map of $^{13}$CO emission.  We have chosen to extract an E-W oriented profile to minimize contamination from the extended emission observed towards the NW of the image, as well as from the southern source.  The FWHM of the observed spatial profile is $\sim1.2''$.  We show that the observed $^{13}$CO spatial profile is well reproduced by our nominal disk model.  

The line emission from AS 205 S, however, is more compact.  Because the zeroth moment $^{12}$CO map is dominated by emission from AS 205 N, we consider in this case a single channel map, shown in Figure \ref{fig:spatial_profile_combined}, which does not suffer any contamination.  In this case, the FWHM of the emission is only $0.9''$ and a model with $R_\mathrm{out}=80$ AU does not fit the data.  Instead, we find that models with $R_\mathrm{out}\sim30-40$ AU provide a reasonable fit.  In principle, the spatial extent of the image also depends on the disk inclination and position angle; however, we find the spatial extent of the model to be relatively insensitive to position angle and inclinations in the range of $15^\circ-35^\circ$.  

This truncation can be compared with expectations from binary disk truncation models.  For example, the parametric fit to truncated disk sizes from \citet{Pichardo05} predicts a relative size ratio for the two disks of a binary that depends only on the binary mass ratio (assuming the disks are close enough to each other to be truncated).  For this system, with $M_1=1.0\, M_\odot$ and $M_2=0.6\, M_\odot$, the predicted ratio of disk sizes would be $\sim1.3$.  Our best-fit disk models have ratios near 2 --- slightly larger than predictions.  With the masses assumed by \citet{Eisner05}, $M_1=1.2\, M_\odot$ and $M_2=1.28\, M_\odot$, the two disks would be expected to be nearly the same size, which is not consistent with our observations.

\subsection{Asymmetric emission and velocity fields}
\subsubsection{Basic characteristics of extended emission}
One of the most striking results of this work is the non-axisymmetric excess emission stretching northeast of AS 205 N, seen at the 10--20$\sigma$ level in the $^{12}$CO zeroth moment map and channel maps (Figures \ref{fig:map_plot}a and \ref{fig:channel_plot_12co}), and 
at the 3$\sigma$ level in the $^{13}$CO zeroth moment map (Figure \ref{fig:map_plot}d).  The $^{12}$CO emission, extending to distances of $\sim2''$, is much wider than the synthesized beam width with major axis $\sim0.7''$ and covers an angular extent of nearly 180$^\circ$.   Figure \ref{fig:channel_plot_12co} also demonstrates that most of the extended emission consists of low velocity gas --- within $\sim$1.2 km s$^{-1}$ of the mean source velocity.  

\subsubsection{Mass and optical depth}
\label{sec:mass}
Using the $^{12}$CO model residuals, we can estimate the mass and optical depth of the spatially extended gas emission. Using the large box shown in Figure \ref{fig:map_plot}c, we find a total flux $1.0\pm0.2$ Jy km s$^{-1}$.  We have also computed the flux without including the negative values, and have tested a range of possible scalings for the disk model (each producing different residuals).  These tests imply that the systematic uncertainty in the total flux is a factor of a few. The temperature of the emission is essentially unknown.  Assuming the emission is optically thin and has a low excitation temperature of 30 K, we find a total mass of $^{12}$CO of $2\times10^{-9}\, M_\odot$, or a total H$_2$ mass of 1$\times10^{-5}\,M_\odot$ assuming an H$_2$/CO ratio of $5000$.  For a much higher excitation temperature of 1000 K, the total mass would be about 20 times larger.  We also compute flux and mass upper limits using the $^{13}$CO residuals in this same box.  We find a total flux of $<0.6$ Jy km s$^{-1}$ (a 3$\sigma$ limit) and a corresponding mass of $<2\times10^{-9} M_\odot$ of $^{13}$CO, implying a $^{12}$CO/$^{13}$CO ratio $>1$.

We can also evaluate the optical depth to determine whether $\tau$ reaches high values.  For the small box in Figure \ref{fig:map_plot}a, assuming the observed flux is proportional to $\tau$ and taking a very conservative (i.e. narrow, to maximize $\tau$) local line broadening of 10 m s$^{-1}$, we compute a CO column density of $\sim10^{-6}$ g cm$^{-2}$, corresponding to an optical depth of 0.9. Therefore, the emission may be optically thick, and we consider our mass estimate a lower limit.  The $^{12}$CO/$^{13}$CO ratio in this box is $>6$, and therefore consistent with either optically thin or moderately optically thick emission, assuming the true isotopic ratio is similar to the values of $\sim50-150$ found in nearby clouds or the interstellar medium \citep{Goto03}.  

\subsubsection{Gas/dust ratio}
Finally, we can estimate the gas/dust ratio of the extended emission, by comparing the line and continuum emission.  In the large box defined in Figure \ref{fig:map_plot}c, there is no detected dust emission above the noise level, which has an RMS value of $\sim2$ mJy beam$^{-1}$. 
Assuming the emission is optically thin, and taking an excitation temperature of 30 K and a grain opacity of $2\,\mathrm{cm^2\, g^{-1}}$ \citep{Beckwith90}, we find a 3$\sigma$ limit to the dust mass of $<5\times10^{-5}\,M_\odot$.  Utilizing the gas mass from above, we find that the gas/dust ratio is $\gtrsim0.4$.  Thus, the constraint on the gas/dust ratio of the extended emission is weak, with both sub-canonical ($<100$) and super-canonical ($>100$) values permitted by the data.

\subsubsection{Non-Keplerian velocity field}
\begin{figure}
\includegraphics[scale=0.4]{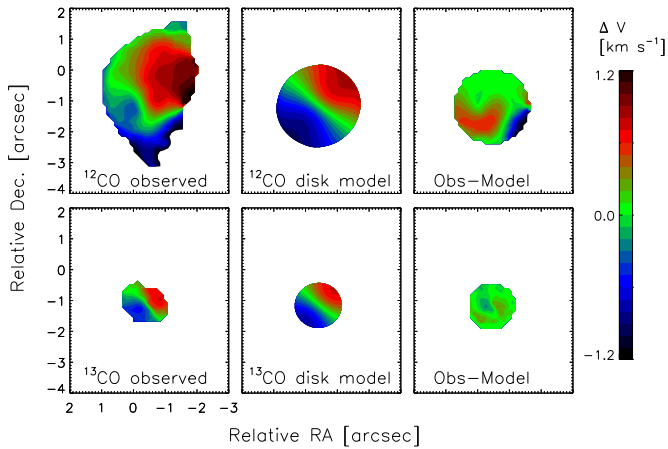}
\caption{Comparison of observed and modeled $^{12}$CO velocity fields (first moment maps), with velocity relative to $v_\mathrm{LSR}= 4.4$ km s$^{-1}$.
\label{fig:mom1_comp_map}}
\end{figure}

The $^{12}$CO and $^{13}$CO first moment maps also show a distinct ``S''-shaped warp in the velocity field, which is different from expectations for a Keplerian disk.  Figure \ref{fig:mom1_comp_map} shows deviations of the observed velocity fields from our nominal disk model.  The observations deviate from expectation by as much as $\sim$1 km s$^{-1}$.

\section{What is the origin of the extended emission and non-Keplerian velocity field?}
\label{sec:origin}
We have identified low-velocity, wide-angle spatially extended $^{12}$CO emission from AS 205 N, as well as non-Keplerian $^{12}$CO and $^{13}$CO velocity fields.  Here we discuss possible physical explanations for this extended and asymmetric emission.  

\subsection{Molecular cloud emission}
\label{section:molecular_cloud}
One possible explanation for our observations is that we are seeing small-scale structure in the surrounding molecular cloud, whose emission is primarily resolved out by the interferometer due to its large spatial scale.  The Ophiuchus cloud region was mapped in $^{12}$CO $1-0$ by \citet{deGeus90}.  AS 205 N lies to the north of most of the prominent filaments in the region, and so is less likely to suffer from cloud contamination than regions closer to the core of Ophiuchus.  More importantly, the local cloud velocity is $\sim$3 km s$^{-1}$ and, as shown in Figure \ref{fig:lineprofile_plot}, this velocity is far from the observed emission line center.  Additionally, much of the spatially extended emission we observe is redder than the emission line center, further separating it from the cloud velocity.   Velocities $\gtrsim$4.5 km s$^{-1}$ are only observed in the $^{12}$CO $1-0$ cloud maps at large distances ($\gtrsim5$ degrees) from AS 205 N.   Therefore, we do not believe that molecular cloud emission can be the cause of the observed spatially extended emission.
  
\subsection{Envelope}
\label{sec:envelope}
Based on a variety of evolutionary indicators, AS 205 N is a Stage II source, with a disk but no envelope.  According to the classification scheme of \citet{Greene94}, Class II objects (believed to represent Stage II) have $2-24\ \mu$m spectral indices, $\alpha=\frac{d\log{\lambda S_\lambda}}{d\log \lambda}$, between $-1.6$ and $-0.3$, while AS 205 N has a 2--24~$\mu$m spectral index in the range of $-1$ to $-0.4$\footnote{The exact numbers depend on whether or not binary-unresolved photometry is included in the calculation.  Photometry from \citet{Eisner05, Liu96, Prato03, McCabe06, Andrews07a, Andrews07b}.}.  AS 205 N also has a bolometric temperature between 1200 and 1400 K$^1$, while Class II objects are defined to have $T_\mathrm{bol}>650$ K \citep{Chen95}.  Furthermore, AS 205 has no detectable envelope in the millimeter maps obtained by \citet{Andre94} (with a beam size of $12''$), while all of the Class 0/I objects in their sample do show resolved envelope emission.  In addition, the total 1.3$\,$mm single dish flux density of 450$\pm90$ mJy/beam (assuming a 20\% flux uncertainty; \citealp{Andre94}) is consistent with our measured system flux density of 441 mJy, suggesting little to no unseen envelope material.  Attributing the 9 mJy difference to an envelope (assuming $T=30 K$ and $\kappa=0.02$ cm$^2$g$^{-1}$, which assumes a gas/dust ratio of 100) would imply a mass of 0.001 $M_\odot$ --- at least an order of magnitude lower than typical envelope masses in Ophiuchus \citep{Andre94}, although the large error bars on the single-dish measurement could allow a mass up to $0.01 \,M_\odot$.  Therefore, AS 205 sits in the Class II regime, although the presence of a very low mass envelope cannot be excluded.

However, these ALMA observations represent a new resolution regime for observations of low-J CO emission from protoplanetary disks, and studies at similar resolution have so far focused on more evolved transition disks (e.g., \citealp{Andrews12}, \citealp{Casassus13}).   Instead, AS 205 N's somewhat high spectral index, as well as its relatively high accretion rate ($7\times10^{-7}\, M_\odot\,$yr$^{-1}$; \citealp{Eisner05}) and disk mass could imply that it sits at the evolutionarily younger end of Stage II.  Therefore, in principle this extended emission could represent the first observation of a small-scale, low-mass remnant of the natal envelope.  However, this hypothesis suffers from stability arguments.    Cloud material at a distance of $2''$  should not be stable in a system with a $1.3''$ binary; this distance is too far to be stable circumprimary material, and too close to be stable circumbinary material \citep{Artymowicz94}.  

\subsection{Accretion from a circumbinary disk}
We consider here whether the observations might be explained as accretion flows from a circumbinary disk.  The arc-like structures seen in the extended $^{12}$CO emission near AS 205 N appear similar to those observed around other $\sim1''$ separation binaries, such as UY Aur, GG Tau and SR 24 \citep{Duvert98, Close98, Roddier96}, structures which are posited to represent accretion streamers from circumbinary disks.  Since this scenario requires the existence of a circumbinary disk, we discuss its possible existence below.

Based on the non-detection of a circumbinary ring in $^{12}$CO emission, AS 205 cannot have a circumbinary disk similar to those around either GG Tau or UY Aur.  GG Tau has a circumbinary disk extending from $\sim1-2''$ from a $0.24''$ binary \citep{White01}, with a total mass of $\sim0.1\, M_\odot$ \citep{Pietu11}.  The UY Aur circumbinary disk has a radius of $\sim$ 500 AU, and a mass of $\sim10^{-3}-10^{-2}\,M_\odot$ \citep{Duvert98}.   Assuming a face-on circumbinary ring extending from $2''-6''$ (250-750 AU), an excitation temperature of 30 K, an H$_2$:CO ratio of 5000, and using a 3 $\sigma$ upper limit derived from our zeroth moment $^{12}$CO map (0.03 Jy km s$^{-1}$ beam$^{-1}$), we derive an upper limit to the mass of a circumbinary disk around AS 205 of $10^{-4}\,M_\odot$ --- significantly lower in mass than that observed around either GG Tau or UY Aur.  Note that we use a CO abundance consistent with the $^{13}$CO abundance utilized by \citet{Duvert98} and so the choice of CO abundance does not affect the comparison between the two disks.  In addition, we note that our observations are sufficiently sensitive to detect disks with similar surface densities as those observed around GG Tau or UY Aur.

The circumbinary disk could potentially remain hidden if it were significantly more extended (and therefore resolved out by the interferometer).  We are sensitive to circumbinary structure up to $\sim 8''$.  Since the observed $^{12}$CO emission extends no more than $\sim2''$, it cannot represent an accretion streamer from an $8''$ wide disk.  Models of this process \citep[e.g.][]{Gunther02, Ochi05, Hanawa10} show continuity of the accretion streamers out to the inner edge of the circumbinary disk.  

It is also possible that significant freeze out of CO could hide a circumbinary disk from our view.  However, single-dish flux measurements for this source place upper limits on the total mass of the system. AS 205 has a single-dish 1.3 mm flux consistent with its interferometric flux (as discussed in Section \ref{sec:envelope}), and therefore consistent with little or no unseen material. Nevertheless, the large error bars on the absolute value of the 1.3mm flux could allow for a mass of up to $\sim0.01\, M_\odot$ (where this implicitly assumes a gas/dus ratio of 100). We note that this situation is in contrast to the binary SR 24, which shows features in scattered light that have been attributed to accretion streamers \citep{Mayama10}, and, therefore, could be considered an analog to AS 205.  SR 24 has a single-dish flux 3--4$\times$ higher than its interferometric flux \citep{Andrews05}, consistent with a significant amount of extended emission ---  a mass of $\sim 0.02-0.03\, M_\odot$, assuming $\kappa_{1.3\,\mathrm{mm}}=0.02\,\mathrm{cm}^2\,\mathrm{g}^{-1}$ and $T=30$ K.  

In conclusion, we show that observations of AS 205 place very firm limits on the possible mass of a CO-rich disk ($\lesssim10^{-4}\, M_\odot$), and the single-dish flux is consistent with no unseen dust, although a total disk mass of up to 0.01 $M_\odot$ (assuming a gas/dust ratio of 100) is allowed by the large error in absolute flux.  With no evidence of a large reservoir to supply mass to the circumstellar disks, we believe it's unlikely that we are observing accretion streams from a circumbinary disk.  Additional single-dish measurements would be helpful to place firmer limits on any possible unseen material in this system.

\subsection{Tidal stripping by AS 205 S}
\label{sec:tidal}
Another possible explanation for our observations is tidal stripping by AS~205 S. 
This explanation has been previously invoked by \citet{Cabrit06} to explain
the morphological and kinematic signatures in the disk around the T Tauri
star RW Auriga A, which bears some interesting similarities to our observations.
RW Aur is a $1.4''$ binary whose $^{12}$CO 2--1 emission shows a red-shifted ``tail''-like feature
extending to $2''$, although at somewhat higher velocities, up to $\sim$7 km s$^{-1}$ --- equivalent to 2--2.5 km s$^{-1}$ if adjusted to the 15$^\circ$ inclination of AS 205 N.

We first evaluate the relevance of tidal effects by computing the tidal radius of
the secondary. For $M_1=1.0\,M_\odot$ and $M_2=0.6\,M_\odot$, the
binary mass ratio is $\mu=M_2/(M_1+M_2)\approx0.38$.  The corresponding Lagrangian
point $L_1$ \citep{Murray99} is located at $\sim0.55r_\mathrm{sep}\sim89$~AU
from $M_1$, i.e., reaching just outside the $\sim$80 AU outer radius of the AS~205 N disk.  
Consequently, some perturbations to the disk symmetry are to be expected.

Tidal effects will be strongest if the encounter is resonant or ``quasi-resonant"
\citep{Donghia10}. Roughly speaking, as in a forced oscillator model,
tidal interactions are most pronounced when there is 
frequency matching between the internal modes of the victim 
(in this case, the disk around AS~205 N) and the external forcing frequency exerted by the perturber, $M_2$. In other words, maximum response occurs at the disk radii $R$ such that the internal orbital frequency $\Omega(R)=\sqrt{GM_1/R^3}$ is an integer fraction of $\Omega_\mathrm{orb}=\sqrt{G(M_1+M_2)/p^3}$, where $p$ is the pericenter distance\footnote
{Strictly speaking, we are looking for an inner Lindblad resonance to
take place within $R_\mathrm{out}$, where the resonance condition
is given by $m(\Omega(R)-s\Omega_\mathrm{orb})\approx\pm\Omega(R)$,
and thus the inner Lindblad resonance occurs at a radius 
$R_\mathrm{ILR}=(1-1/m)^{2/3}(1-\mu)^{1/3}p$~.
}.  Note that the resonance also requires a prograde orbit of the perturber with respect to the victim's spin.  

A single passage by the perturber can excite prominent two-armed spirals (similar in morphology to what we observe) in the victim
provided the resonance occurs within the victim's disk's radial extent \citep[see, e.g.,][]{Barnes96, Mihos96,Donghia10}.
However, if the pericenter distance, $p$, is equal to the binary projected separation of $\sim$160 AU or larger, low order near-resonance is not achieved for radii inside the primary's disk with our assumed $\mu=0.38$.  To achieve low-order resonance, AS 205 S would need to be more massive.   As discussed in Section \ref{sec:models}, there is some uncertainty about the mass of AS 205 S.  With the AS~205 S mass estimate derived by \citet{Eisner05}, $\sim1.3\,M_\odot$,  resonant forcing would occur at $\sim70$~AU, just inside the outer radius of the AS~205 N disk.  

However, even if AS 205 S is massive enough to have low-order resonant interactions with the disk of AS 205 N, a major drawback of this explanation is that bound pairs are expected to be already devoid of gas at the lowest order resonances that create strong two-armed features, since those resonances would be quickly cleared out after a few subsequent pericenter passages \citep{Paczynski77,Artymowicz94}.  Therefore, we would need to be observing AS~205 during its first, or nearly first, encounter.  This is unlikely if AS 205 is a bound pair, as the orbital period is $\sim$ a few thousand years, while the estimated age of AS 205 is $10^5$ years \citep{Prato03}.  

It is also very unlikely that this is a first (flyby) encounter between two stars. \citet{Thies05} suggest that close stellar encounters may be likely during the $\sim$6 Myr lifetime of disks around stars in massive young clusters; with encounter probabilities of 10-30\% for encounter distances of 60--200 AU (assuming a cluster mass of $500\, M_\odot$ and a size of 0.3 pc), one can expect $\sim$100 close stellar encounters throughout the lifetime of the cluster.  However, Ophiuchus is significantly less massive and dense than this model presumes, with a stellar mass of $\sim53\, M_\odot$ in the central 2 pc$^2$ of the $\rho$ Oph cloud core \citep{Kenyon98}, and, furthermore, AS 205 lies in a less dense region at the edge of the Ophiuchus cloud \citep{deGeus90}.  Assuming $53\, M_\odot$ in 2 pc$^2$, and using the equations found in \citeauthor{Thies05}, we find encounter probabilities of a few $\times10^{-5}$ out to 300 AU --- equivalent to an expected number of collisions of a few $\times10^{-3}$.  Finally, we must compute not the likelihood of an encounter, but the likelihood of observing an encounter in progress.  If we assume an encounter velocity of 1 km s$^{-1}$ and an impact parameter of 500 AU, the encounter lasts $\sim$2000 years ---  a factor of 500 shorter than the cluster age of at least 1 Myr \citep[e.g.][]{Wilking05}; therefore, the probability of viewing an encounter is 500 times lower than the probability of an encounter occurring.  These factors combined suggest that the probability of observing a flyby encounter with an impact parameter of a few 100 AU in this region is $< 10^{-5}$.

Although lower-order resonances do not appear to explain the features we observe, in the long run, the
evolution of circumprimary and circumbinary disks is dominated by higher-order, weaker resonances. 
As with the lower-order resonances, the resonant forcing and the internal damping of the disk reach a quasi-stationary state,
in which the disk size is nearly in equilibrium. A balance is reached between the
viscous and the tidal torques, eventually defining the truncation radius of both circumprimary
and circumsecondary disks \citep{Artymowicz94}.  However, one subtlety in the analysis by 
\citet{Artymowicz94} is that the equilibrium truncation radius is reached only after a viscous relaxation timescale that depends on the 
binary eccentricity, the final disk radius, the disk aspect ratio and the disk viscosity. Therefore, 
for very young systems, transient responses on the disks are still possible during every pericenter passage, especially for 
very eccentric binaries.   Unfortunately, at this point we are not able to pursue this analysis further
due to the largely unconstrained orbital properties of this system, and the uncertain mass of AS 205 S. 

\subsection{Winds}

Our observations might also be explained by some form of wind.  Thermal (photoevaporative) winds are unlikely to explain our observations, as the driving radiation for these flows tends to destroy molecules \citep[e.g.][]{Gorti09a}.  On the other hand, magneto-hydrodynamic (MHD) disk winds, which are believed to be (at least partly) responsible for the larger-scale bipolar jets and outflows emanating from young stars \citep[e.g.][and references therein]{Ferreira06, Pudritz07, Ray07}, can produce relatively low-velocity winds with a significant molecular content \citep{Panoglou12}.  

MHD disk winds have launching velocities that are similar to the sound speed \citep[e.g.][]{Kurosawa06}, and, therefore, tracers of the wind launching region can have relatively low velocities.  Temperatures in the launch region range from a few 1000 K in the inner disk, down to $<100$ K in the outer disk, with associated velocities in the range of $\sim$1--3 km s$^{-1}$.  Observed launching velocities will be even lower if the streamlines are not directed towards the observer.   However, poloidal velocities increase with distance along a streamline, reaching velocities as high as $\sim$ a few 100 km s$^{-1}$ in existing models  \citep{Pudritz07}, which are much higher than we observe here.  Another obstacle for disk wind models may be whether there is sufficient overall gas density to explain the extended emission.  In the models of \citet{Panoglou12}, CO densities of $\sim0.1\ \mathrm{cm}^{-3}$ result in a theoretical column density of $\sim10^{-9}\ \mathrm{g\ cm}^{-2}$, assuming a column with vertical height 20 AU.  This is three orders of magnitude below the observed lower limit to the column density, $\sim10^{-6}\ \mathrm{g\ cm}^{-2}$.  However, if lower terminal velocities are possible, then mass conservation will correspondingly increase the density in the wind, allowing models to match observations.  We consider this in Section \ref{sec:windmodels}.

\subsubsection{Relationship to prior observations of outflows}
The structure observed here is significantly different from the very collimated millimeter ouflows observed around Class 0 and Class I disks \citep[e.g.][]{Jorgensen07}, many of which additionally appear to be tracing shocks on the inner surface of an outflow cavity in a surrounding envelope.  However, \citet{Arce06} imaged a sample of nine Class 0, I and II young stellar objects in $^{12}$CO 1-0 and noted that the millimeter emission becomes less collimated with evolutionary state.  In fact, the emission we observe bears some resemblance to the structures observed by \citeauthor{Arce06} near the Class II sources T Tau and GK/GI Tau. The size scales of these structure are larger (up to $\sim40''$ in length), but our observations are only sensitive to structures up to $\sim8''$ in size.  The velocities of these outflows are also higher than what we observe, typically 2--5 km s$^{-1}$ offset from the source velocity.  In general, however, millimeter observations of class II sources with outflows are limited, and even fewer observations have the spatial resolution of our observations, making comparisons difficult.  

AS 205 N has some spectroscopic evidence for an active outflow.  Although to our knowledge no optical jets have yet been imaged from this source, \citet{Mundt84} reported strong, narrow, blueshifted ($\sim150$ km s$^{-1}$) absorption features in Na D lines that are consistent with outflowing gas and observed in other sources with known outflows.  In addition, \citet{Rigliaco13} report a high-velocity component in observations of the [O I] 6300 \AA\ line from AS 205 N, indicative of an outflow.  It is also important to note that our observations were {\it motivated} by the suggestion of \citet{Pontoppidan11} that a disk wind may explain observed single-peaked rovibrational CO emission lines, as well as asymmetric spectro-astrometric profiles of these same lines.  In the next section, we test whether the model utilized by \citeauthor{Pontoppidan11} can also explain our observations.

\subsubsection{Comparison with model predictions}
\label{sec:windmodels}
A simple parameterization of a disk wind was implemented by \citet{Pontoppidan11} to fit observed rovibrational emission lines, and incorporated into the radiative transfer code RADLite \citep{Pontoppidan09}.  We describe this model and our choice of nominal model parameters in detail in the Appendix.

\begin{figure*}[!ht]
\includegraphics[scale=0.5]{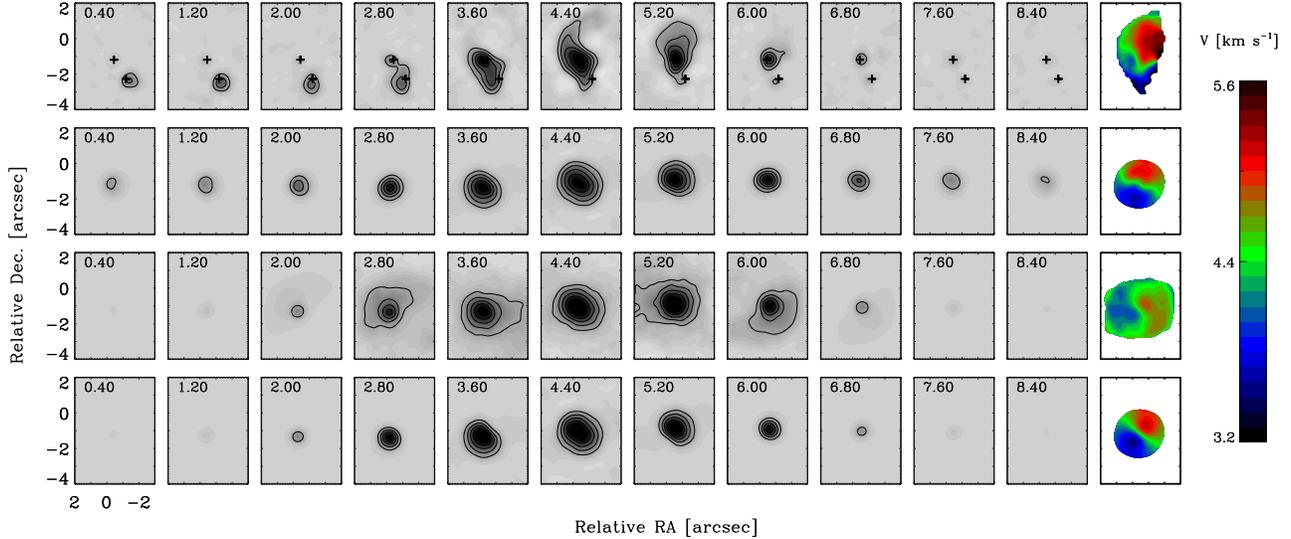}
\caption{$^{12}$CO channel maps and first moment maps for, from top to bottom: observed data, our nominal disk+wind model, a disk+wind model with lower terminal velocity and higher mass-loss rate, and a disk-only model.  Note the extended velocity coverage compared to Figure \ref{fig:channel_plot_12co}.
\label{fig:channel_plot_12co_43_30_54_large}}
\end{figure*}

\begin{figure}[!ht]
\includegraphics[scale=0.4]{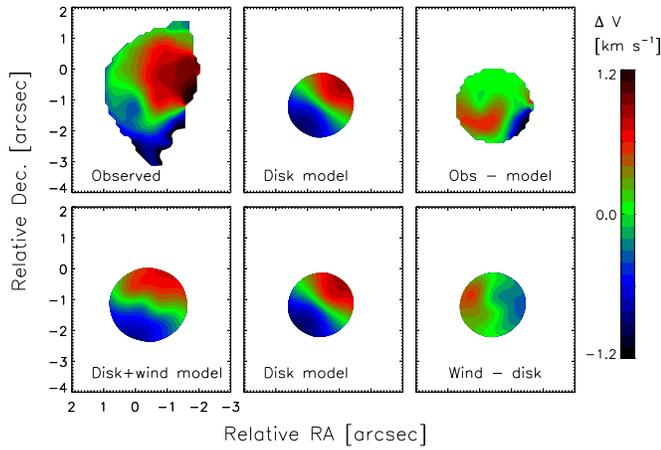}
\caption{Top: Observed and disk model $^{12}$CO first moment maps, relative to $v_\mathrm{LSR}= 4.4$ km s$^{-1}$, as well as their difference.  Bottom: Disk+wind and disk-only models as well as their difference.  All models assume the far side of the disk is located towards the northeast. \label{fig:mom1_comp_map_wind}}
\end{figure}

We find that the nominal wind model does not produce any appreciable extended emission, beyond that expected for a disk, due to the low predicted densities at distances of 100's of AU from the central star.   However, the model does reproduce some of the other features of our observations.  As shown in Figure \ref{fig:channel_plot_12co_43_30_54_large}, it produces additional flux at higher velocities compared to a disk-only model.  The wind model also produces deviations from a Keplerian velocity field, including a curved first moment map, demonstrated in Figure \ref{fig:mom1_comp_map_wind}.  The curvature in the model is due to two effects.  At small radii, the dominant effect of the wind is that it makes the azimuthal velocities sub-Keplerian.  At larger radii, the poloidal velocities are larger and begin to dominate the differences, causing the far side of the disk (in this case, towards the northeast) to be preferentially redshifted and the near side (towards the southwest) to be preferentially blueshifted.   There are some similarities between the modeled and observed first moment maps, in that both show a complex ``S''-like shape, although it is clear that the observed moment map is more complex than the modeled map.  

A twisted velocity field can also arise from radial inflow, as suggested by \citet{Rosenfeld14} to explain the velocity field of the transitional disk around HD 142527.  Inflow and outflow can only be distinguished if the near and far sides of the disks are identified.  In detail, HD 142527 is not an analog to AS 205 N; the fast radial inflow in HD 142527 is proposed to reconcile its high accretion rate and inner clearing, while AS 205 N has a full disk with no inner clearing.  In addition, radial inflow alone would not explain the very extended $^{12}$CO emission we observe.  However, some wind models themselves predict that accretion may be dominated by restricted regions of the disk producing fast flows \citep{Bai13}.  The resultant signature in the velocity field would then depend on the relative contributions of inflowing and outflowing gas to the observed emission, and could vary throughout the disk, creating a complex velocity field.  Unfortunately, disk-wide simulations of this process do not yet exist for comparison with observations.

Although the {\it nominal} wind model does not produce any appreciable extended emission, the density of the wind depends on the assumed mass-loss rate and terminal velocity.  Figure \ref{fig:channel_plot_12co_43_30_54_large} shows predicted channel maps for an alternative wind model, with a low terminal velocity (2 km s$^{-1}$) and high mass-loss rate ($2\times10^{-7} M_\odot$ yr$^{-1}$), which shows appreciable extended emission.  The detailed structure does not obviously match our observations.  It should be noted, however, that dynamical timescales for the binary orbit and the wind are similar, and so AS 205 S should cause perturbations to the wind structure.  Therefore, our observations may be explained by a {\it combination} of disk winds and binary interaction.

In conclusion, wind models are in some ways consistent with the observations of AS 205 N.  In particular, models produce a curved first moment map with some similarities with the observed $^{12}$CO velocity field.   However, existing wind models do not readily produce significant extended emission.  Extended emission instead requires that the models be modified to have lower terminal velocities and/or higher mass-loss rates.  Although the geometry of the resulting emission does not precisely match our observations, the detailed features we observe cannot be produced by an axisymmetric model.  They could, however, 
be produced by interactions between the wind and the stellar companion, AS 205 S.

\section{DISCUSSION}
\label{sec:discussion}
\subsection{Why hasn't this extended emission been observed before?}
These are the highest resolution observations of rotational CO emission from AS 205 to date.  AS 205 was observed in $^{12}$CO $3-2$ by \citet{Andrews09} with a beam size of $2.19''\times 1.60''$ and in $^{12}$CO $2-1$ by \citet{Oberg11a} with a beam size of $3.1'' \times 2.0''$.  In both cases, the beam size was not sufficient even to resolve the binary.  Since the extended structure we observe is only a few arcsec in size, a resolution of $\sim1''$ or better is required.  

Only a few disks have been observed in molecular lines at a resolution of $\sim1''$ or better to date, as pre-ALMA observations were limited to the brightest disks. Therefore, it is difficult to say whether the phenomenon we observe is a common or unusual feature for protoplanetary disks.  Interestingly, molecular line imaging at or near this resolution often seems to reveal surprises --- for example, an extended CO disk around TW Hya \citep{Andrews12}, an asymmetric, perhaps tidally-induced ``arm'' near RW Aur \citep{Cabrit06} and asymmetric spiral arms in the disk of AB Aur \citep{Corder05, Pietu05}.  ALMA observations of a few disks have already revealed azimuthal asymmetries and non-Keplerian flows \citep{Casassus13, Christiaens14,Rosenfeld14}, as well as a helical disk wind \citep{Klaassen13}.  Therefore, it seems likely that the resolution and sensitivity afforded by ALMA will continue to reveal many new small-scale asymmetric structures.

If the extended structure we observe here is a common feature for protoplanetary disks, this could have implications for the interpretation of lower-resolution observations.  For low resolution observations, the observed CO flux could incorrectly be attributed solely to a disk.  Although it's difficult to determine the precise contribution to the flux from the wind, the box shown in Figure \ref{fig:map_plot}c contributes 5\% of the total observed $^{12}$CO $2-1$ line flux.  Also, if the wind has a different temperature than the disk, interpretation of observations of multiple CO isotopologues could incorrectly determine the gas temperatures.  Since gas temperatures are in turn used to determine gas masses \citep[e.g.][]{Panic08}, this could have important consequences for our understanding of the planet-forming capabilities of disks.  

\subsection{If we are observing mass loss, what is the estimated mass-loss rate?}
If we are actually observing a wind, with some simple assumptions, we can estimate the mass-loss rate.  For this calculation, we use the CO observed in the large box in Figure \ref{fig:map_plot}c (see Section \ref{sec:mass}) and multiply by 4 to account for all directions of the wind.  We obtain $M_\mathrm{gas}\sim8\times10^{-5}\, M_\odot$ for an excitation temperature of 30 K, or $M_\mathrm{gas}\sim2\times10^{-3}\,M_\odot$ for 1000 K.  The mass-loss rate is $\dot{M}_\mathrm{loss}\approx M_\mathrm{gas} v_\mathrm{gas}/D_\mathrm{gas}$ where $v_\mathrm{gas}$ and $D_\mathrm{gas}$ are the actual velocity and length of the outflow.  Both $D_\mathrm{gas}$ and $v_\mathrm{gas}$ are different from the observed distance and velocity, where $D_\mathrm{obs}=D_\mathrm{gas}\sin{\theta}$, $v_\mathrm{obs}=v_\mathrm{gas}\cos{\theta}$ and $\theta$ is the angle between the outflow and the observer.  Thus, $\dot{M}_\mathrm{loss}\approx \frac{M_\mathrm{gas} v_\mathrm{obs}\sin{\theta}}{D_\mathrm{obs}\cos{\theta}}$.  Taking $D_\mathrm{obs}\sim300$ AU, $v_\mathrm{obs}\sim1\,$ km s$^{-1}$ and assuming $\theta=15^\circ$ (equal to the best-fit disk inclination), we find $\dot{M}_\mathrm{loss}\sim 10^{-8} - 10^{-7} M_\odot \mathrm{yr}^{-1}$.  A different assumption about the inclination of the flow can either raise (for $i>15^\circ$) or lower (for $i<15^\circ$) the mass-loss rate.  The disk + wind models described in Section \ref{sec:windmodels} also required a rate of $\sim10^{-7} M_\odot \mathrm{yr}^{-1}$ to produce significant extended emission.

Using observations of CO rovibrational lineshapes and spectro-astrometry along with disk+wind radiative transfer models, \citet{Pontoppidan11} estimated a similar or slightly lower mass-loss rate for AS 205 N --- $9\times10^{-9}\, M_\odot\, \mathrm{yr}^{-1}$.  Both results therefore suggest that $\dot{M}_\mathrm{loss}/\dot{M}_\mathrm{acc}\sim0.01-0.1$.  

\section{CONCLUSIONS}
\label{sec:conclusions}
We have presented here one of the most sensitive and highest resolution mm-wave studies of a nearby T Tauri binary system and its accompanying gas disks, a study which was enabled by the newly available ALMA facility.  It is clear from these observations that AS 205 N does not solely consist of a Keplerian gas disk, with differences including extended asymmetric $^{12}$CO emission and a non-Keplerian velocity field.  Although we have examined here several possible scenarios to explain these observations, no single scenario is able to completely explain these features.  Two of the most promising explanations include a low-velocity wind, or tidal stripping by the companion, but potential models will need to be explored further and fine-tuned if they are to match the observations.  It may also be that we are observing wind-driven mass-loss, modified by tidal interactions with the secondary.  Future ALMA observations of disks in close binary systems as well as disks with IR spectroscopic evidence for disk winds, or evidence for larger-scale outflows, will be helpful for distinguishing between the two possible explanations.

Aside from the observations presented here, it is known that AS 205 N has other somewhat unusual properties compared to typical T Tauri disks.  However, it is {\it not} unique, as there appears to be an emerging sub-population of disks sharing properties with AS 205 N, properties including high accretion rates and veiling, as well as strongly single-peaked IR emission line profiles \citep[e.g.][]{Gahm08, Bast11, Pontoppidan11}.  The story that these disks have to tell us is still unclear, but we are excited about the prospects of continuing to explore their interesting properties, and expect that the mysteries of their origin will soon be revealed.

\section{Acknowledgements}
We would like to thank the staff at the North American ALMA Science Center, especially Scott Schnee, for their help with the data reduction.  C.S. would like to acknowledge helpful discussions with Ilaria Pascucci, Sean Andrews and Katherine Rosenfeld.  C.S. would also like to acknowledge the financial support of the NOAO Leo Goldberg Fellowship program.  D.M. would like to acknowledge helpful discussions with Dong Lai and Robert Harris.  This paper makes use of the following ALMA data: ADS/JAO.ALMA\#2011.0.00531.S. ALMA is a partnership of ESO (representing its member states), NSF (USA) and NINS (Japan), together with NRC (Canada) and NSC and ASIAA (Taiwan), in cooperation with the Republic of Chile. The Joint ALMA Observatory is operated by ESO, AUI/NRAO and NAOJ.  The National Radio Astronomy Observatory is a facility of the National Science Foundation operated under cooperative agreement by Associated Universities, Inc.

\appendix

\section{RADLite implementation of wind model}
The wind model used in this work is based on the split-monopole parameterization described by \citet{Kurosawa06}, and originally from \citet{Knigge95}, in which the poloidal velocities follow a $\beta$-velocity law, the toroidal velocities conserve angular momentum, and the density is determined via mass conservation.  The equations for velocity and density as a function of cylindrical radius, $w$, and distance along the streamline, $l$, are given in \citet{Kurosawa06}.  The geometry of this model is completely defined by the (virtual) launching point, which is located a distance, $d$, below the star, and inner and outer radial extents, $w_{\mathrm{wi}}$ and $w_{\mathrm{wo}}$.  This parameterized model is meant to approximate magnetocentrifugal wind models, originally suggested by \citet{Blandford82} and \citet{Pudritz83}.

The wind model was implemented into the line radiative transfer code RADLite by \citet{Pontoppidan11}.  RADLite takes as input an arbitrary density, velocity and temperature grid.  In this implementation the densities and velocities above some defined surface level, $\tau_\mathrm{surf}$, are simply replaced by the equations found in \citet{Kurosawa06}.  $\tau_\mathrm{surf}$ is a vertical optical depth computed at 0.55 $\mu$m. The temperatures, and the densities below $\tau_\mathrm{surf}$, take whatever form the user may choose; in our case, they are described by the disk model discussed in Section \ref{sec:models}.  Parameters used in our nominal model are similar to those used by \citeauthor{Pontoppidan11} and are listed in Table \ref{table:parameters}.   Short definitions of the parameters are provided; detailed information about these parameters can be found in \citet{Kurosawa06}.

We also note a few clarifications.  In our implementation, we replace the product $v_\mathrm{esc}\times f$, the product of the local escape velocity and the ratio of the terminal velocity to the escape velocity, with a single parameter, $v_\mathrm{term}$, representing the terminal velocity of the wind.  Although $v_\mathrm{esc}$ varies with radius, in practice the small launching region of the wind precludes a large range in $v_\mathrm{esc}$, so this substitution is a useful conceptual simplification.  In addition, we provide the total mass-loss rate ($\dot{M}$), integrated over the disk surface participating in the wind, although we also include the mass-loss rate per unit area at the innermost launching point ($\dot{m}$) defined in \citet{Kurosawa06}.

\begin{deluxetable}{lll}
\tablecaption{Disk wind parameters\label{table:parameters}}
\tablehead{\colhead{Parameter} & \colhead{Value} &Notes}
\startdata
d & 18.5 $R_\odot$ &1\\
$\beta$&2&2\\
$v_\mathrm{term}$&20 km s$^{-1}$&3\\
$\dot{M}_\mathrm{wind}$ & $1.5\times10^{-9}\ M_\odot\ \mathrm{yr}^{-1}$&4\\
$\dot{m}_\mathrm{wind}$ & $2.8\times10^{-9}\ \mathrm{g}\ \mathrm{cm}^{-2} \mathrm{s}^{-1}$&5\\
$p$ &-7/2&6\\
$R_s$&19 AU&7\\ 
$\tau_\mathrm{surf}$&0.75&8\\
$w_{\mathrm{wi}}$&0.14 AU&9\\
$w_{\mathrm{wo}}$&1 AU&10\\
\enddata
\tablenotetext{1}{Distance between the star and the wind locus point, equivalent to $5 R_\star$ for AS 205 N, where $R_\star=3.7 R_\odot$.}
\tablenotetext{2}{Wind acceleration parameter}
\tablenotetext{3}{Wind terminal velocity}
\tablenotetext{4}{Total wind mass-loss rate}
\tablenotetext{5}{Mass-loss rate per unit area, at the innermost wind launching radius}
\tablenotetext{6}{Power-law exponent for radial dependence of mass-loss rate}
\tablenotetext{7}{Wind scale length}
\tablenotetext{8}{$\tau$ at 0.55$\mu$m taken as the launching height of the wind}
\tablenotetext{9}{Cylindrical radius of the innermost wind launching radius, here taken to be the inner edge of the disk.}
\tablenotetext{10}{Cylindrical radius of the outermost wind launching radius}
\end{deluxetable}


\begin{thebibliography}{}
\bibitem[Alencar et al.(2001)]{Alencar01} Alencar, S.~H.~P., 
Johns-Krull, C.~M., \& Basri, G.\ 2001, \aj, 122, 3335 
\bibitem[Andrews 
\& Williams(2005)]{Andrews05} Andrews, S.~M., \& Williams, J.~P.\ 2005, \apjl, 619, L175 
\bibitem[Andrews \& Williams(2007b)]{Andrews07b} Andrews, S.~M., \& Williams, J.~P.\ 2007, \apj, 671, 1800 
\bibitem[Andrews \& Williams(2007a)]{Andrews07a} Andrews, S.~M., \& Williams, J.~P.\ 2007, \apj, 659, 705 
\bibitem[Andrews et al.(2009)]{Andrews09} Andrews, S.~M., Wilner, 
D.~J., Hughes, A.~M., Qi, C., \& Dullemond, C.~P.\ 2009, \apj, 700, 1502 
\bibitem[Andrews et al.(2010)]{Andrews10} Andrews, S.~M., Wilner, 
D.~J., Hughes, A.~M., Qi, C., \& Dullemond, C.~P.\ 2010, \apj, 723, 1241 
\bibitem[Andrews et al.(2012)]{Andrews12} Andrews, S.~M., Wilner, 
D.~J., Hughes, A.~M., et al.\ 2012, \apj, 744, 162 
\bibitem[Andre 
\& Montmerle(1994)]{Andre94} Andre, P., \& Montmerle, T.\ 1994, \apj, 420, 837 
\bibitem[Arce \& Sargent(2006)]{Arce06} Arce, H.~G., \& Sargent, A.~I.\ 2006, \apj, 646, 1070 
\bibitem[Artymowicz 
\& Lubow(1994)]{Artymowicz94} Artymowicz, P., \& Lubow, S.~H.\ 1994, \apj, 421, 651 
\bibitem[Artymowicz 
\& Lubow(1996)]{Artymowicz96} Artymowicz, P., \& Lubow, S.~H.\ 1996, \apjl, 467, L77 
\bibitem[Bai 
\& Stone(2013)]{Bai13} Bai, X.-N., \& Stone, J.~M.\ 2013, \apj, 769, 76 
\bibitem[Barnes 
\& Hernquist(1996)]{Barnes96} Barnes, J.~E., \& Hernquist, L.\ 1996, \apj, 471, 115 
\bibitem[Bast et al.(2011)]{Bast11} Bast, J.~E., Brown, J.~M., Herczeg, G.~J., van Dishoeck, E.~F., \& Pontoppidan, K.~M.\ 2011, \aap, 527, A119 
\bibitem[Beckwith et al.(1990)]{Beckwith90} Beckwith, S.~V.~W., 
Sargent, A.~I., Chini, R.~S., \& Guesten, R.\ 1990, \aj, 99, 924 
\bibitem[Blandford 
\& Payne(1982)]{Blandford82} Blandford, R.~D., \& Payne, D.~G.\ 1982, \mnras, 199, 883 
\bibitem[Brown et al.(2013)]{Brown13} Brown, J.~M., 
Pontoppidan, K.~M., van Dishoeck, E.~F., et al.\ 2013, \apj, 770, 94 
\bibitem[Bruderer et 
al.(2012)]{Bruderer12} Bruderer, S., van Dishoeck, E.~F., Doty, S.~D., \& Herczeg, G.~J.\ 2012, \aap, 541, A91 
\bibitem[Bruderer et al.(2014)]{Bruderer14} Bruderer, S., van der Marel, N., van Dishoeck, E.~F., \& van Kempen, T.~A.\ 2014, \aap, 562, A26 
\bibitem[Cabrit et 
al.(2006)]{Cabrit06} Cabrit, S., Pety, J., Pesenti, N., \& Dougados, C.\ 2006, \aap, 452, 897 
\bibitem[Casassus et al.(2013)]{Casassus13} Casassus, S., van der 
Plas, G., M, S.~P., et al.\ 2013, \nat, 493, 191 
\bibitem[Chen et al.(1995)]{Chen95} Chen, H., Myers, P.~C., 
Ladd, E.~F., \& Wood, D.~O.~S.\ 1995, \apj, 445, 377 
\bibitem[Christiaens et al.(2014)]{Christiaens14} Christiaens, V., 
Casassus, S., Perez, S., van der Plas, G., 
\& M{\'e}nard, F.\ 2014, \apjl, 785, L12 
\bibitem[Close et al.(1998)]{Close98} Close, L.~M., Dutrey, A., 
Roddier, F., et al.\ 1998, \apj, 499, 883 
\bibitem[Corder et al.(2005)]{Corder05} Corder, S., Eisner, J., 
\& Sargent, A.\ 2005, \apjl, 622, L133 
\bibitem[D'Onghia et al.(2010)]{Donghia10} D'Onghia, E., 
Vogelsberger, M., Faucher-Giguere, C.-A., 
\& Hernquist, L.\ 2010, \apj, 725, 353 
\bibitem[Dullemond 
\& Dominik(2004)]{Dullemond04} Dullemond, C.~P., \& Dominik, C.\ 2004, \aap, 417, 159  
\bibitem[Dutrey et al.(1997)]{Dutrey97} Dutrey, A., Guilloteau, S., \& Guelin, M.\ 1997, \aap, 317, L55 
\bibitem[Duvert et 
al.(1998)]{Duvert98} Duvert, G., Dutrey, A., Guilloteau, S., et al.\ 1998, \aap, 332, 867 
\bibitem[Eisner et al.(2005)]{Eisner05} Eisner, J.~A., 
Hillenbrand, L.~A., White, R.~J., Akeson, R.~L., 
\& Sargent, A.~I.\ 2005, \apj, 623, 952 
\bibitem[Ferreira et 
al.(2006)]{Ferreira06} Ferreira, J., Dougados, C., \& Cabrit, S.\ 2006, \aap, 453, 785 
\bibitem[de Geus et 
al.(1990)]{deGeus90} de Geus, E.~J., Bronfman, L., \& Thaddeus, P.\ 1990, \aap, 231, 137 
\bibitem[Gahm et 
al.(2008)]{Gahm08} Gahm, G.~F., Walter, F.~M., Stempels, H.~C., Petrov, P.~P., \& Herczeg, G.~J.\ 2008, \aap, 482, L35 
\bibitem[Ghez et al.(1993)]{Ghez93} Ghez, A.~M., Neugebauer, 
G., \& Matthews, K.\ 1993, \aj, 106, 2005 
\bibitem[Gorti 
\& Hollenbach(2009)]{Gorti09a} Gorti, U., \& Hollenbach, D.\ 2009, \apj, 690, 1539 
\bibitem[Goto et al.(2003)]{Goto03} Goto, M., Usuda, T., Takato, N., et al.\ 2003, \apj, 598, 1038 
\bibitem[Greene et al.(1994)]{Greene94} Greene, T.~P., Wilking, 
B.~A., Andre, P., Young, E.~T., \& Lada, C.~J.\ 1994, \apj, 434, 614 
\bibitem[G{\"u}nther 
\& Kley(2002)]{Gunther02} G{\"u}nther, R., \& Kley, W.\ 2002, \aap, 387, 550 
\bibitem[Hanawa et al.(2010)]{Hanawa10} Hanawa, T., Ochi, Y., 
\& Ando, K.\ 2010, \apj, 708, 485 
\bibitem[Herczeg et al.(2005)]{Herczeg05} Herczeg, G.~J., Walter, 
F.~M., Linsky, J.~L., et al.\ 2005, \aj, 129, 2777 
\bibitem[J{\o}rgensen et al.(2007)]{Jorgensen07} J{\o}rgensen, 
J.~K., Bourke, T.~L., Myers, P.~C., et al.\ 2007, \apj, 659, 479 
\bibitem[Kenyon et al.(1998)]{Kenyon98} Kenyon, S.~J., Lada, 
E.~A., \& Barsony, M.\ 1998, \aj, 115, 252 
\bibitem[Klaassen et 
al.(2013)]{Klaassen13} Klaassen, P.~D., Juhasz, A., Mathews, G.~S., et al.\ 2013, \aap, 555, A73 
\bibitem[Knigge et al.(1995)]{Knigge95} Knigge, C., Woods, 
J.~A., \& Drew, J.~E.\ 1995, \mnras, 273, 225 
\bibitem[Koresko(2002)]{Koresko02} Koresko, C.~D.\ 2002, \aj, 124, 1082 
\bibitem[Kurosawa et al.(2006)]{Kurosawa06} Kurosawa, R., Harries, 
T.~J., \& Symington, N.~H.\ 2006, \mnras, 370, 580 
\bibitem[Lacy et al.(1994)]{Lacy94} Lacy, J.~H., Knacke, R., 
Geballe, T.~R., \& Tokunaga, A.~T.\ 1994, \apjl, 428, L69 
\bibitem[Liu et al.(1996)]{Liu96} Liu, M.~C., Graham, J.~R., 
Ghez, A.~M., et al.\ 1996, \apj, 461, 334 
\bibitem[Loinard et al.(2008)]{Loinard08} Loinard, L., Torres, 
R.~M., Mioduszewski, A.~J., 
\& Rodr{\'{\i}}guez, L.~F.\ 2008, \apjl, 675, L29 
\bibitem[Mamajek(2008)]{Mamajek08} Mamajek, E.~E.\ 2008, 
Astronomische Nachrichten, 329, 10 
\bibitem[Mayama et al.(2010)]{Mayama10} Mayama, S., Tamura, M., 
Hanawa, T., et al.\ 2010, Science, 327, 306 
\bibitem[McCabe et al.(2006)]{McCabe06} McCabe, C., Ghez, A.~M., 
Prato, L., et al.\ 2006, \apj, 636, 932 
\bibitem[McMullin et al.(2007)]{McMullin07} McMullin, J.~P., 
Waters, B., Schiebel, D., Young, W., 
\& Golap, K.\ 2007, Astronomical Data Analysis Software and Systems XVI, 376, 127
\bibitem[Melo(2003)]{Melo03} Melo, C.~H.~F.\ 2003, \aap, 410, 269 
\bibitem[Mihos 
\& Hernquist(1996)]{Mihos96} Mihos, J.~C., \& Hernquist, L.\ 1996, \apj, 464, 641 
\bibitem[Mundt(1984)]{Mundt84} Mundt, R.\ 1984, \apj, 280, 749  
\bibitem[Munoz\&Kratter(2014)]{Munoz14} Munoz, D.\ \& Kratter, K., in prep
\bibitem[Murray 
\& Dermott(1999)]{Murray99} Murray, C.~D., \& Dermott, S.~F.\ 1999, Solar system dynamics by Murray, C.~D., 1999,  
\bibitem[Najita et al.(2003)]{Najita03} Najita, J., Carr, J.~S., 
\& Mathieu, R.~D.\ 2003, \apj, 589, 931 
\bibitem[Natta et 
al.(2006)]{Natta06} Natta, A., Testi, L., \& Randich, S.\ 2006, \aap, 452, 245 
\bibitem[{\"O}berg et al.(2010)]{Oberg10} {\"O}berg, K.~I., Qi, 
C., Fogel, J.~K.~J., et al.\ 2010, \apj, 720, 480 
\bibitem[{\"O}berg et al.(2011)]{Oberg11a} {\"O}berg, K.~I., Qi, 
C., Fogel, J.~K.~J., et al.\ 2011, \apj, 734, 98 
\bibitem[Ochi et al.(2005)]{Ochi05} Ochi, Y., Sugimoto, K., 
\& Hanawa, T.\ 2005, \apj, 623, 922 
\bibitem[Paczynski(1977)]{Paczynski77} Paczynski, B.\ 1977, \apj, 
216, 822 
\bibitem[Pani{\'c} et 
al.(2008)]{Panic08} Pani{\'c}, O., Hogerheijde, M.~R., Wilner, D., \& Qi, C.\ 2008, \aap, 491, 219 
\bibitem[Panoglou et 
al.(2012)]{Panoglou12} Panoglou, D., Cabrit, S., Pineau Des For{\^e}ts, G., et al.\ 2012, \aap, 538, A2 
\bibitem[Pichardo et al.(2005)]{Pichardo05} Pichardo, B., Sparke, 
L.~S., \& Aguilar, L.~A.\ 2005, \mnras, 359, 521 
\bibitem[Pi{\'e}tu et 
al.(2005)]{Pietu05} Pi{\'e}tu, V., Guilloteau, S., \& Dutrey, A.\ 2005, \aap, 443, 945 
\bibitem[Pi{\'e}tu et 
al.(2007)]{Pietu07} Pi{\'e}tu, V., Dutrey, A., \& Guilloteau, S.\ 2007, \aap, 467, 163 
\bibitem[Pi{\'e}tu et 
al.(2011)]{Pietu11} Pi{\'e}tu, V., Gueth, F., Hily-Blant, P., Schuster, K.-F., \& Pety, J.\ 2011, \aap, 528, A81 
\bibitem[Pontoppidan et al.(2011)]{Pontoppidan11} Pontoppidan, K.~M., 
Blake, G.~A., \& Smette, A.\ 2011, \apj, 733, 84 
\bibitem[Pontoppidan et al.(2009)]{Pontoppidan09} Pontoppidan, K.~M., 
Meijerink, R., Dullemond, C.~P., \& Blake, G.~A.\ 2009, \apj, 704, 1482 
\bibitem[Pontoppidan et al.(2010)]{Pontoppidan10b} Pontoppidan, K.~M., 
Salyk, C., Blake, G.~A., et al.\ 2010, \apj, 720, 887 
\bibitem[Prato et al.(2003)]{Prato03} Prato, L., Greene, T.~P., 
\& Simon, M.\ 2003, \apj, 584, 853 
\bibitem[Pudritz 
\& Norman(1983)]{Pudritz83} Pudritz, R.~E., \& Norman, C.~A.\ 1983, \apj, 274, 677 
\bibitem[Pudritz et al.(2007)]{Pudritz07} Pudritz, R.~E., Ouyed, 
R., Fendt, C., \& Brandenburg, A.\ 2007, Protostars and Planets V, 277 
\bibitem[Ray et al.(2007)]{Ray07} Ray, T., Dougados, C., 
Bacciotti, F., Eisl{\"o}ffel, J., 
\& Chrysostomou, A.\ 2007, Protostars and Planets V, 231 
\bibitem[Reipurth 
\& Zinnecker(1993)]{Reipurth93} Reipurth, B., \& Zinnecker, H.\ 1993, \aap, 278, 81 
\bibitem[Rigliaco et al.(2013)]{Rigliaco13} Rigliaco, E., 
Pascucci, I., Gorti, U., Edwards, S., 
\& Hollenbach, D.\ 2013, \apj, 772, 60 
\bibitem[Roddier et al.(1996)]{Roddier96} Roddier, C., Roddier, 
F., Northcott, M.~J., Graves, J.~E., \& Jim, K.\ 1996, \apj, 463, 326 
\bibitem[Rosenfeld et al.(2012)]{Rosenfeld12} Rosenfeld, K.~A., Qi, 
C., Andrews, S.~M., et al.\ 2012, \apj, 757, 129 
\bibitem[Rosenfeld et al.(2013)]{Rosenfeld13} Rosenfeld, K.~A., 
Andrews, S.~M., Hughes, A.~M., Wilner, D.~J., 
\& Qi, C.\ 2013, \apj, 774, 16 
\bibitem[Rosenfeld et al.(2014)]{Rosenfeld14} Rosenfeld, K.~A., 
Chiang, E., \& Andrews, S.~M.\ 2014, \apj, 782, 62 
\bibitem[Salyk et al.(2013)]{Salyk13} Salyk, C., Herczeg, 
G.~J., Brown, J.~M., et al.\ 2013, \apj, 769, 21 
\bibitem[Simon et al.(2000)]{Simon00} Simon, M., Dutrey, A., 
\& Guilloteau, S.\ 2000, \apj, 545, 1034 
\bibitem[Salyk et al.(2008)]{Salyk08} Salyk, C., Pontoppidan, 
K.~M., Blake, G.~A., et al.\ 2008, \apjl, 676, L49 
\bibitem[Thies et al.(2005)]{Thies05} Thies, I., Kroupa, P., 
\& Theis, C.\ 2005, \mnras, 364, 961 
\bibitem[Wang et al.(2004)]{Wang04} Wang, H., Mundt, R., 
Henning, T., \& Apai, D.\ 2004, \apj, 617, 1191 
\bibitem[White 
\& Ghez(2001)]{White01} White, R.~J., \& Ghez, A.~M.\ 2001, \apj, 556, 265 
\bibitem[Wilking et al.(2005)]{Wilking05} Wilking, B.~A., Meyer, 
M.~R., Robinson, J.~G., \& Greene, T.~P.\ 2005, \aj, 130, 1733 
\bibitem[Yang et al.(2012)]{Yang12} Yang, H., Herczeg, G.~J., 
Linsky, J.~L., et al.\ 2012, \apj, 744, 121 
\bibitem[Zhang et al.(2013)]{Zhang13} Zhang, K., Pontoppidan, 
K.~M., Salyk, C., \& Blake, G.~A.\ 2013, \apj, 766, 82 
\end{thebibliography}
\end{document}